\documentclass{osa-article}

\journal{oe}
\articletype{Research Article}
\begin{document}

\title{Enhancement of photon blockade effect via quantum interference}

\author{Fen Zou\authormark{1}, Deng-Gao Lai\authormark{1}, and Jie-Qiao Liao\authormark{1,$\dagger$}}

\address{\authormark{1}Key Laboratory of Low-Dimensional Quantum Structures and Quantum Control of
Ministry of Education, Key Laboratory for Matter Microstructure and Function of Hunan Province, Department of Physics and Synergetic Innovation Center for Quantum Effects and Applications, Hunan Normal University, Changsha 410081, China}

\email{\authormark{$\dagger$}jqliao@hunnu.edu.cn} %% email address is required

\begin{abstract}
We study the photon blockade effect in a coupled cavity system, which is formed by a linear cavity coupled to a Kerr-type nonlinear cavity via a photon-hopping interaction. We explain the physical phenomenon from the viewpoint of the conventional and unconventional photon blockade effects. The corresponding physical mechanisms of the two kinds of photon blockade effects are based on the anharmonicity in the eigenenergy spectrum and the destructive quantum interference between two different transition paths, respectively. In particular, we find that the photon blockade via destructive quantum interference also exists in the conventional photon blockade regime, and that the unconventional photon blockade occurs in both the weak- and strong-Kerr nonlinearity cases. The photon blockade effect can be observed by calculating the second-order correlation function of the cavity field. This model is general and hence it can be implemented in various experimental setups such as coupled optical-cavity systems, coupled photon-magnon systems, and coupled superconducting-resonator systems. We present some discussions on the experimental feasibility.
\end{abstract}

\section{Introduction}

The photon blockade effect describes a physical phenomenon that the occupation of one photon in a cavity will block the consequent injection of the second photon~\cite{imamoglu1997Strongly,carusotto2013Quantum}. The basic idea of photon blockade is in analogy to the concept of the Coulomb blockade~\cite{fulton1987Observation} of electrons in mesoscopic physics: the electrons inside an island will create a strong coulomb repulsion preventing other electrons to flow. Conventionally, photonic nonlinearity in the eigenenergy spectrum is considered as the physical origin of the occurrence of photon blockade~\cite{imamoglu1997Strongly}. Owing to the anharmonicity of the eigenenergy spectrum, a resonant physical transition between the low-excitation states will lead to an off-resonance in the consequent transition to the upper energy levels, and hence the injection of the second photon is blockaded. In contrast to the conventional photon blockade~\cite{imamoglu1997Strongly}, a new kind of physical mechanism for creating photon blockade was recently proposed. This new physical mechanism (called as unconventional photon blockade) is based on the destructive quantum interference between different excitation pathways~\cite{liew2010Single,bamba2011Origin}. As a result, the unconventional photon blockade mechanism usually requires that the physical systems involve several degrees of freedom so that multi-transition paths can be established.

So far, both of the above mentioned physical mechanisms have been widely studied in various quantum systems. The conventional photon blockade effect has been studied in quantum optical systems such as cavity quantum electrodynamics (QED) systems~\cite{tian1992Quantum,birnbaum2005Photon,faraon2008Coherent,peyronel2012Quantum,reinhard2012Strongly,ridolfo2012Photon,
bajcsy2013Photon,miranowicz2014State,muller2015Coherent,deng2015Enhancement,poshakinskiy2016Biexcitonmediated,zhu2017Collective,radulaski2017Photon,
hamsen2017TwoPhoton,han2018Electromagnetic,bin2018Twophoton,trivedi2019Photon,kowalewska2019Two} and circuit-QED systems~\cite{hoffman2011Dispersive,lang2011Observation,bourassa2012Josephsonjunctionembedded,liu2014Blockade}, the Kerr-type nonlinear cavities~\cite{imamoglu1997Strongly,ferretti2010Photon,liao2010Correlated,miranowicz2013Twophoton,hovsepyan2014Multiphoton,
huang2018Nonreciprocal,ghosh2019Dynamicala}, optomechanical systems~\cite{rabl2011Photon,ludwig2012Enhanced,
stannigel2012Optomechanical,liao2013Photon,liao2013Correlated,xu2013Photoninduced,kronwald2013Full,
lu2013Quantumcriticalityinduced,komar2013Singlephoton,wang2015Tunable,hu2015Quantum,zhu2018Controllable,zou2019Enhancement}, the doubly resonant nonlinear cavity systems\cite{majumdar2013Single}, and atomic systems~\cite{huang2013Photon}. Meanwhile, the unconventional photon blockade effect has been studied in coupled Kerr-cavity systems~\cite{bamba2011Origin,xu2014Strong,shen2015Tunable,flayac2015Allsilicon,flayac2017Unconventional}, the cavity-QED systems~\cite{tang2015Quantum}, coupled optomechanical systems~\cite{xu2013Antibunching,gerace2014Unconventional,sarma2018Unconventional,li2019Nonreciprocal}, and other optical systems~\cite{flayac2013Inputoutput,lemonde2014Antibunching,zhou2015Unconventional,sarma2017Quantuminterferenceassisted}. In these schemes, either the light-matter interactions or the moving boundary of the fields has been introduced to manipulate the optical mode so that the photon blockade effect can be realized. In particular, the unconventional photon blockade effect has recently been realized experimentally in a quantum dot cavity
system~\cite{snijders2018Observation} and in a superconducting circuit consisting
of two coupled resonators~\cite{vaneph2018Observation}. An analogous phenomenon of phonon blockade has also been predicted in various quantum systems~\cite{liu2010Qubitinduced,wang2016Method,xie2017Phonon,xu2016Phonon}.

In this paper, we study the photon blockade effect in a coupled cavity system consisting of a linear cavity and a Kerr-type nonlinear cavity. Here the two cavities are coupled with each other through a photon-hopping interaction. Our study is inspired by the recent theoretical advances in coupled optical-cavity systems~\cite{gerace2009Quantumoptical}, and experimental advances in coupled photon-magnon systems~\cite{wang2016Magnon,wang2018Bistability}, a hybrid system formed by a microwave resonator and a phononic resonator~\cite{arrangoiz-arriola2018Coupling}, and coupled superconducting-resonator systems~\cite{eichler2014QuantumLimited} formed by a linear resonator and a Kerr-type nonlinear resonator~\cite{kirchmair2013Observation}. These experimental systems can be described by a common physical model considered in this paper. In addition, the study of the statistical properties of cavity photons in coupled nonlinear cavities has attracted much attention~\cite{hartmann2010Polariton,bardyn2012Majoranalike,jin2013Photon}. Therefore, it is natural to consider a simpler model consisting of both linear and nonlinear cavities. For simplicity and without loss of generality, below we use the terms of optical cavity and all the results work for other physical degrees of freedom. The photon blockade effect is studied by calculating the equal-time second-order correlation function of the linear cavity mode. We analyze how does the nonlinear cavity mode modulate the linear cavity mode through two different physical mechanisms corresponding to conventional and unconventional photon blockade effects. In particular, we investigate the anharmonicity of the eigenenergy spectrum and destructive quantum interference effect between two transition paths in this system. By analyzing the relationship between the anharmonicity of the eigenenergy spectrum and the linewidth of the energy levels, we study the inherent parameter condition under which the conventional photon blockade can be realized. Interestingly, we find that there exists quantum interference effect in the conventional photon blockade regime in this system. We also derive the parameter condition under which the unconventional photon blockade occurs. It is found that the unconventional photon blockade based on destructive quantum interference between two different transition paths not only exists in the weak-nonlinearity regime~\cite{liew2010Single,bamba2011Origin}, but also can take place in the strong-nonlinearity regime. In addition, we find some symmetric relations in the parameter conditions of the unconventional photon blockade.

\section{Physical system}

\begin{figure}
\center
\includegraphics[width=12cm]{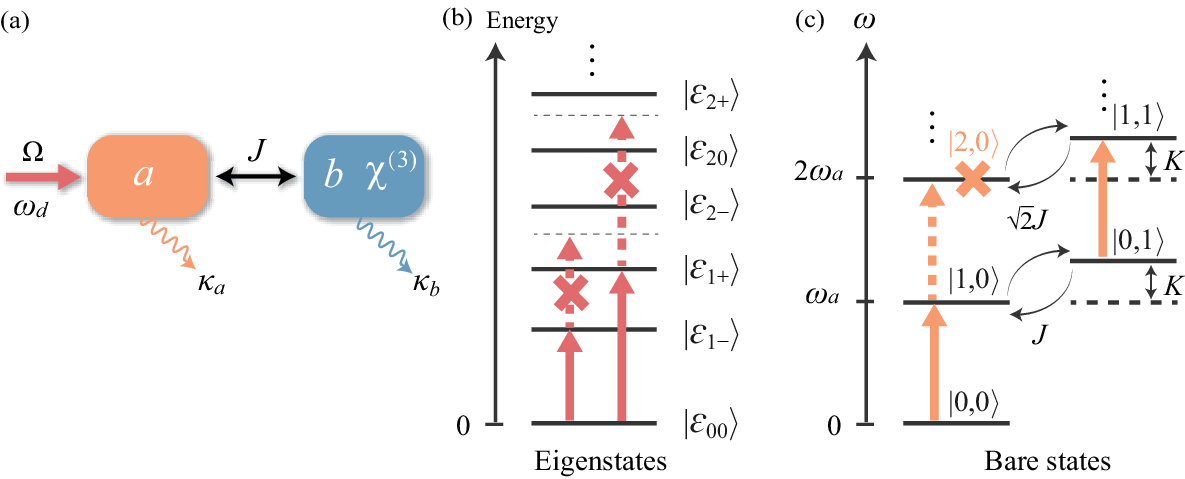} %bb=1 4 275 258,
\caption{(a) Schematic diagram of the coupled cavity model composed of a linear cavity coupled to a Kerr-type nonlinear cavity through a photon-hopping interaction. (b) The eigenenergy spectrum of the coupled cavity system in the subspace with zero, one, and two excitations. (c) Energy-level diagram of the bare states of the system in the low-excitation subspace.}
\label{Fig1}
\end{figure}

We consider a coupled cavity system, which is formed by a linear cavity coupled to a Kerr-type nonlinear cavity through a photon-hopping interaction [Fig.~\ref{Fig1}(a)]. In order to manipulate the quantum state of this coupled cavity system, a monochromatic field is introduced to drive the linear cavity. The Hamiltonian of the total system reads ($\hbar=1$)
\begin{equation}
H_{\text{sys}}=\omega_{a}a^{\dagger}a+\omega_{b}b^{\dagger}b+Kb^{\dagger}bb^{\dagger}b+J(a^{\dagger}b+b^{\dagger}a)+\Omega(a^{\dagger}e^{-i\omega_{d}t}+ae^{i\omega_{d}t}),\label{eq1}
\end{equation}
where $a$ ($a^{\dagger}$) and $b$ ($b^{\dagger}$) are, respectively, the annihilation (creation) operators of the two cavity modes, with the corresponding resonance frequencies $\omega_{a}$ and $\omega_{b}$. The term $Kb^{\dagger}bb^{\dagger}b$ represents the Kerr nonlinearity in the nonlinear cavity, with $K$ being the Kerr nonlinear coefficient. The parameter $J$ denotes the strength of the photon-hopping interaction between the two cavity modes. The parameters $\Omega$ and $\omega_{d}$ are the driving amplitude and driving frequency of the linear cavity, respectively. In a rotating frame with respect to $H_{0}=\omega_{d}(a^{\dagger}a+b^{\dagger}b)$, the Hamiltonian of the system becomes
\begin{equation}
H_{I}=\Delta_{a}a^{\dagger}a+\Delta_{b}b^{\dagger}b+Kb^{\dagger}bb^{\dagger}b+J(a^{\dagger}b+b^{\dagger}a)+\Omega(a^{\dagger}+a),  \label{Hinterpic}
\end{equation}
where we introduce the driving detunings $\Delta_{a}=\omega_{a}-\omega_{d}$ and $\Delta_{b}=\omega_{b}-\omega_{d}$.

In the absence of the cavity field driving, the Hamiltonian of the coupled cavity system reads $H_{\text{ccs}}=\Delta_{a}a^{\dagger}a+\Delta_{b}b^{\dagger}b+Kb^{\dagger}bb^{\dagger}b+J(a^{\dagger}b+b^{\dagger}a)$, and the total excitation number
operator $\hat{N}=a^{\dagger}a+b^{\dagger}b$ is a conserved quantity in this system because of the commutative relation $[H_{\text{ccs}},\hat{N}]=0$. To study the photon blockade effect in this system, we consider the weak-driving regime. In this case, the photon number involved is small and hence we can restrict the system within the low-excitation subspaces. Below, we will consider the total excitation numbers as $N=0$, $1$, and $2$ in our analytical studies. The subspaces corresponding to $N=0$, $1$, and $2$ are spanned over the basis states $\{|0,0\rangle\}$, $\{|1,0\rangle,|0,1\rangle\}$, and $\{|2,0\rangle,|1,1\rangle,|0,2\rangle\}$, respectively. Here $\vert m,n\rangle$ represents the state with $m$ photons in the linear cavity and $n$ photons in the nonlinear cavity. The eigensystems of the Hamiltonian $H_{\text{ccs}}$ in these three subspaces can be obtained as follows. In the zero-excitation subspace, the eigen-equation is $H_{\text{ccs}}|\varepsilon_{00}\rangle=E_{00}|\varepsilon_{00}\rangle$ with the eigenstate $|\varepsilon_{00}\rangle=|0,0\rangle$ and the eigenvalue $E_{00}=0$. In the single-excitation subspace, the eigen-equation is $H_{\text{ccs}}|\varepsilon_{1\pm}\rangle=E_{1\pm}|\varepsilon_{1\pm}\rangle$, where the eigenstates and eigenvalues are defined by
\begin{equation}
|\varepsilon_{1\pm}\rangle=C_{0,1}^{[1\pm]}|0,1\rangle+C_{1,0}^{[1\pm]}|1,0\rangle,
\end{equation}
and
\begin{equation}
E_{1\pm}=\frac{\Delta_{a}+\Delta_{b}+K}{2}\pm\frac{\sqrt{(\Delta_{b}-\Delta_{a}+K)^{2}+4J^{2}}}{2}.
\end{equation}
The superposition coefficients in the eigenstates are defined by
\begin{equation}
C_{0,1}^{[1+]}=C_{1,0}^{[1-]}=\cos\theta,\hspace{0.3 cm}C_{1,0}^{[1+]}=-C_{0,1}^{[1-]}=\sin\theta,
\end{equation}
where the mixing angle $\theta$ is defined by $\tan(2\theta)=2J/(\Delta_{b}-\Delta_{a}+K)$.
%%%%%%%%%%%%%%%%%%%%%
%%%%%%%%%%%%%%%%%%%%%%

In the two-excitation subspace, the eigenstates and eigenvalues can be obtained by solving the eigensystem of the matrix
\begin{equation}
H_{\text{ccs}}^{[2]}=\left(
\begin{array}{ccc}
2\Delta_{b}+4K & \sqrt{2}J & 0 \\
\sqrt{2}J & \Delta_{a}+\Delta_{b}+K & \sqrt{2}J \\
0 & \sqrt{2}J & 2\Delta_{a}
\end{array}\right),\label{Hamilin2excspace}
\end{equation}
which is defined based on the basis states $|0,2\rangle=(1,0,0)^{T}$, $|1,1\rangle=(0,1,0)^{T}$, and $|2,0\rangle=(0,0,1)^{T}$, where ``$T$'' denotes the matrix transpose. The eigensystem
of the Hamiltonian in the two-excitation subspace is defined by $H_{\text{ccs}}^{[2]}|\varepsilon_{2s}\rangle=E_{2s}|\varepsilon_{2s}\rangle$ with $s=\pm,0$.
The eigenvalues are given by
\begin{align}
E_{2-}&=-\frac{1}{3}\{A+\sqrt{-3D}[\cos(\alpha/3)+\sqrt{3}\sin(\alpha/3)]\},  \notag \\
E_{20}&=-\frac{1}{3}\{A+\sqrt{-3D}[\cos(\alpha/3)-\sqrt{3}\sin(\alpha/3)]\},  \notag \\
E_{2+}&=-\frac{1}{3}[A-2\sqrt{-3D}\cos(\alpha/3)],
\end{align}
with the corresponding eigenstates
\begin{equation}
|\varepsilon_{2s}\rangle=C_{0,2}^{[2s]}|0,2\rangle+C_{1,1}^{[2s]}|1,1\rangle+C_{2,0}^{[2s]}|2,0\rangle\label{eigenstain2excspace}
\end{equation}
for $s=\pm,0$. The superposition coefficients and the relating parameters used in the eigensystem are defined by
\begin{align}
C_{0,2}^{[2s]}&=-\sqrt{2}J(E_{2s}-2\Delta_{a}) N_{2s}^{-1/2},\quad
C_{2,0}^{[2s]}=\sqrt{2}J(2\Delta_{b}+4K-E_{2s})N_{2s}^{-1/2},\notag\\
C_{1,1}^{[2s]}&=(E_{2s}-2\Delta_{a})(2\Delta_{b}+4K-E_{2s})N_{2s}^{-1/2},
\end{align}
and
\begin{align}
A &=-5K-3\Delta_{a}-3\Delta_{b},\notag\\
B &=-4J^{2}+4K^{2}+14K\Delta_{a}+2\Delta_{a}^{2}+6K\Delta_{b}+8\Delta_{a}\Delta_{b}+2\Delta_{b}^{2},  \notag\\
C &=8J^{2}K+4J^{2}\Delta_{a}-8K^{2}\Delta_{a}-8K\Delta_{a}^{2}+4J^{2}\Delta_{b}-12K\Delta_{a}\Delta_{b}
-4\Delta_{a}^{2}\Delta_{b}-4\Delta_{a}\Delta_{b}^{2},  \notag\\
D &=B-\frac{1}{3}A^{2},  \quad
E =C+\frac{2}{27}A^{3}-\frac{1}{3}AB,  \quad
\alpha =\arccos [-3E\sqrt{-3D}/(2D^{2})],  \notag\\
N_{2s} &=(E_{2s}-2\Delta_{a})^{2}[2J^{2}+(2\Delta_{b}+4K-E_{2s})^{2}]+2J^{2}(2\Delta_{b}+4K-E_{2s})^{2}.
\end{align}

In Fig.~\ref{Fig1}(b), we show the energy spectrum of $\hat{H}_{\text{ccs}}$ in these subspaces associated with zero, one, and two photons. The transitions are induced by driving the cavity under the single-photon resonance condition. The off-resonance in the second-photon transition, which is induced by the anharmonicity in the eigenenergy spectrum, is the physical origin of the conventional photon blockade effect. In Fig.~\ref{Fig1}(c), we show the picture of quantum interference effect determining the unconventional photon blockade. The quantum interference effect occurs between these two paths~\cite{bamba2011Origin}: (i) the direct excitation from $\vert 1,0\rangle\overset{\Omega}{\rightarrow}\vert 2,0\rangle $ and (ii) the indirect transition path $\vert 1,0\rangle \overset{J}{\leftrightarrow}\vert 0,1\rangle\overset{\Omega}{\rightarrow}\vert1,1\rangle \overset{J}{\leftrightarrow}\vert 2,0\rangle$. The unconventional photon blockade is achieved based on the destructive quantum interference between two different transition paths of bare states. In the coupled-cavity system, the eigenstates of the system are the superposition of these bare states $\vert m,n\rangle$ in the same subspace, and then the bare states can be expressed by the superposition of these eigenstates. Hence, the occupation of $\vert2,0\rangle$ contains some cross terms, which are induced by quantum interference among these eigenstates $\vert \varepsilon_{2s} \rangle$ ($s=\pm,0$). Particularly, the unconventional photon blockade via the destructive quantum interference can be observed when the occupation of $\vert2,0\rangle$ is zero. In fact, the descriptions of the two kinds of quantum interference are essentially identical, and the difference between these two descriptions is the analysis of quantum interference effect (one is from the eigenstates and the other one is from the bare states). Note that the electromagnetically induced transparency, which is a result of the Fano interference among different transition pathways, has an analogous quantum interference effect~\cite{peng2014What}.

\section{Photon blockade effect}

The photon blockade effect in the linear cavity can be characterized by calculating the equal-time second-order correlation function
\begin{equation}
g_{a}^{(2)}(0)\equiv\frac{\langle a^{\dagger}a^{\dagger}aa\rangle_{\mathrm{ss}}}{\langle a^{\dagger}a\rangle_{\mathrm{ss}}^{2}},
\end{equation}
where the average value is taken over the steady state of the system. The correlation function can be calculated numerically~\cite{johansson2012QuTiP,johansson2013QuTiP} by solving the quantum master equation
\begin{equation}
\dot{\rho}=i[\rho,H_{I}]+\frac{\kappa_{a}}{2}(\bar{n}_{a}+1)\mathcal{L}_{a}[\rho]
+\frac{\kappa_{a}}{2}\bar{n}_{a}\mathcal{L}_{a^{\dagger}}[\rho]
+\frac{\kappa_{b}}{2}(\bar{n}_{b}+1)\mathcal{L}_{b}[\rho]+\frac{\kappa_{b}}{2}\bar{n}_{b}\mathcal{L}_{b^{\dagger}}[\rho],  \label{masteq}
\end{equation}
where the Hamiltonian $H_{I}$ is given in Eq.~(\ref{Hinterpic}) and $\mathcal{L}_{o}[\rho]=(2o\rho o^{\dagger}-o^{\dagger}o\rho-\rho o^{\dagger}o)$ denotes the Lindblad superoperator for an operator $o$~\cite{scully_zubairy_1997}. The parameters $\kappa_{a}$ and $\kappa_{b}$ are, respectively, the dissipation rates of the two cavity fields, and $\bar{n}_{a}=[\exp(\hbar\omega_{a}/k_{B}T_{a})-1]^{-1}$ and $\bar{n}_{b}=[\exp(\hbar\omega_{b}/k_{B}T_{b})-1]^{-1}$ are the average thermal excitation numbers of the baths at temperatures $T_{a}$ and $T_{b}$, with $k_{B}$ being the Boltzmann constant.

For simplicity, we consider the case of zero-temperature environments, and then we can use the effective Hamiltonian method to describe the evolution of the system. Here the evolution of the system is governed by the non-Hermitian Hamiltonian which is formed by adding phenomenologically the imaginary dissipation terms into Hamiltonian~(\ref{Hinterpic}) as follows~\cite{Carmichael1993}
\begin{equation}
H_{\text{nHmt}}=(\Delta_{a}-i\kappa_{a}/2)a^{\dagger}a+(\Delta_{b}-i\kappa_{b}/2)b^{\dagger}b
+Kb^{\dagger}bb^{\dagger}b+J(a^{\dagger}b+b^{\dagger}a)+\Omega(a^{\dagger}+a).\label{Hnonheimite}
\end{equation}
In the weak-driving case ($\Omega/\kappa_{a}\ll1$), we expand the wave function of the system with the bare-state bases as
\begin{equation}
\vert\psi\rangle=C_{0,0}(t)\vert 0,0\rangle+C_{1,0}(t)\vert 1,0\rangle+C_{0,1}(t)\vert 0,1\rangle+C_{2,0}(t)\vert 2,0\rangle+C_{1,1}(t)\vert 1,1\rangle+C_{0,2}(t)\vert 0,2\rangle,\label{psi}
\end{equation}
where $C_{m,n}(t)$ for $m,n=0,1,2$ are the probability amplitudes corresponding to the bare state $\vert m,n\rangle$. Based on the Hamiltonian in Eq.~(\ref{Hnonheimite}) and the wave function in Eq.~(\ref{psi}), we can obtain the equations of motion for these probability amplitudes $C_{m,n}$ as
\begin{align}
i\dot{C}_{0,0}(t) &=\Omega C_{1,0}(t),  \nonumber \\
i\dot{C}_{1,0}(t) &=(\Delta_{a}-i\kappa_{a}/2)C_{1,0}(t)+JC_{0,1}(t)+\Omega C_{0,0}(t)
+\sqrt{2}\Omega C_{2,0}(t),  \nonumber \\
i\dot{C}_{0,1}(t) &=JC_{1,0}(t)+(\Delta_{b}-i\kappa_{b}/2)C_{0,1}(t)+KC_{0,1}(t)
+\Omega C_{1,1}(t),  \nonumber \\
i\dot{C}_{2,0}(t) &=(2\Delta_{a}-i\kappa_{a})C_{2,0}(t)+\sqrt{2}JC_{1,1}(t)+\sqrt{2}\Omega C_{1,0}(t),  \nonumber \\
i\dot{C}_{1,1}(t) &=\Omega C_{0,1}(t)+\sqrt{2}JC_{2,0}(t)+[\Delta _{a}+\Delta_{b}-i(\kappa_{a}+\kappa_{b})/2]C_{1,1}(t)+KC_{1,1}(t)+\sqrt{2}JC_{0,2}(t),  \nonumber \\
i\dot{C}_{0,2}(t) &=\sqrt{2}JC_{1,1}(t)+(2\Delta_{b}-i\kappa_{b})C_{0,2}(t)+4KC_{0,2}(t).
\end{align}
In the weak-driving case, these probability amplitudes can also be classified into various groups of different orders of the small ratio $\Omega/\kappa_{a}$. The amplitude $C_{0,0}(t)$ is of the zero order of $\Omega/\kappa_{a}$. The coefficients $C_{1,0}(t)$ and $C_{0,1}(t)$ are of the first order of $\Omega/\kappa_{a}$, and $C_{2,0}(t)$, $C_{1,1}(t)$, and $C_{0,2}(t)$ are of the second order of $\Omega/\kappa_{a}$. In this case, we can solve the equations of motion for the probability amplitudes $C_{m,n}(t)$ using the perturbation method~\cite{carmichael1991Quantum}, namely discarding the higher-order terms in the equations of motion for the lower-order variables. Below, we consider the case of $\Delta_{a}=\Delta_{b}=\Delta$ and $\kappa_{a}=\kappa_{b}=\kappa$, and then the steady-state solutions of the probability amplitudes can be obtained by setting $\partial C_{m,n}/\partial t=0$ as
\begin{align}
C_{0,0} &=1,\notag \\
C_{1,0} &=2(2K-i\kappa+2\Delta)\Omega/[4J^{2}+(\kappa+2i\Delta)(2iK+\kappa+2i\Delta)],\notag \\
C_{0,1} &=-4J\Omega/[4J^{2}+(\kappa+2i\Delta)(2iK+\kappa+2i\Delta)],\notag \\
C_{2,0} &=2\sqrt{2}[(2K-i\kappa+2\Delta)(4K-i\kappa+2\Delta)(K-i\kappa+2\Delta)+4J^{2}K]\Omega^{2}M^{-1},  \notag \\
C_{1,1} &=-8J(4K-i\kappa +2\Delta)(K-i\kappa+2\Delta) \Omega ^{2}M^{-1},\notag \\
C_{0,2} &=8\sqrt{2}J^{2}(K-i\kappa+2\Delta)\Omega^{2}M^{-1},\label{stssoluprobaamplit}
\end{align}
where $M$ is defined as
\begin{align}
M &=[4J^{2}+(\kappa+2i\Delta)(2iK+\kappa+2i\Delta)][(\kappa+2i\Delta)(4K-i\kappa+2\Delta)(iK+\kappa+2i\Delta)\notag\\
&\quad+4J^{2}(2K-i\kappa+2\Delta)].
\end{align}

Then the equal-time second-order correlation function of the linear cavity can be approximately expressed as
\begin{equation}
g_{a}^{(2)}(0)=\frac{2P_{|2,0\rangle}}{(P_{|1,0\rangle}+P_{|1,1\rangle}+2P_{|2,0\rangle})^{2}}\approx\frac{2P_{|2,0\rangle}}{P_{|1,0\rangle}^{2}},  \label{gensolution}
\end{equation}
where the probabilities of finding $m$ and $n$ photons respectively in the linear and nonlinear cavities are given by $P_{|m,n\rangle}=|C_{m,n}|^{2}$. The condition $g_{a}^{(2)}(0)<1$ $[g_{a}^{(2)}(0)>1]$ indicates the sub-Poissonian (super-Poissonian) photon statistics, and the correlation function $g_{a}^{(2)}(0)\ll1$ is a signature of the photon blockade effect.

\section{Results}

\subsection{Conventional photon blockade}

\begin{figure}[tbp]
\center
\includegraphics[width=9cm]{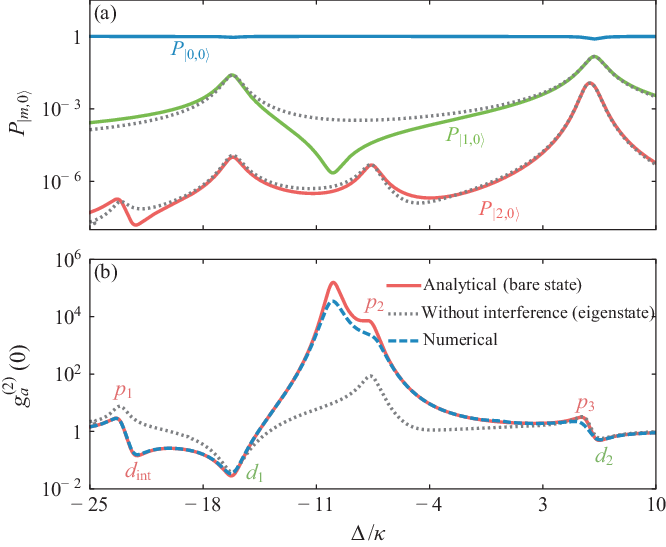}
\caption{(a) The state occupations $P_{|m,0\rangle}$ for the bare states $|m,0\rangle$ ($m=0,1,2$) versus the scaled driving detuning $\Delta/\kappa$. The colored solid curves and the gray dotted curves are plotted based on the exact results and the non-quantum-interference part of the exact results, respectively. (b) The equal-time second-order correlation function $g_{a}^{(2)}(0)$ versus the driving detuning $\Delta/\kappa$. The blue dashed curve is plotted using the numerical solution of Eq.~(\ref{masteq}), while the red solid (gray dotted) curve corresponds to the exact (non-quantum-interference) analytical result. Other parameters are $\kappa_{a}=\kappa_{b}=\kappa$, $\Omega/\kappa=0.3$, $K/\kappa=10$,  $J/\kappa=10$, and $\bar{n}_{a}=\bar{n}_{b}=0$.}
\label{Fig2}
\end{figure}

As we know, the conventional photon blockade can be explained from the viewpoint of the anharmonicity in the eigenenergy spectrum. Usually, the photon blockade effect is determined by the transition resonance, and hence the driving detuning is an important parameter to control the occurrence of photon blockade. To see how does the driving detuning affect the photon number distributions, in Fig.~\ref{Fig2}(a) we plot the state occupations $P_{|m,0\rangle}$ for the bare states $|m,0\rangle$ ($m=0,1,2$) versus the scaled driving detuning $\Delta/\kappa$. We can see that the probability distributions satisfy the relation $P_{|0,0\rangle}\gg P_{|1,0\rangle}\gg P_{|2,0\rangle}$ due to the weak driving. The state occupation $P_{|0,0\rangle}$ is almost one, which means that almost all the probability of the system is in state $|0,0\rangle$. Moreover, we can see that there are two peaks for the state occupation $P_{|1,0\rangle}$ (the green solid curve). To understand the locations of the peaks in $P_{|1,0\rangle}$, we need to analyze the eigenenergy spectrum of the Hamiltonian $H_{\text{ccs}}$ and the resonance condition in these transitions. In this system, there are two transition channels for the single photon transitions $|\varepsilon_{00}\rangle\rightarrow |\varepsilon_{1\pm}\rangle$. As a result, the single photon resonance conditions are determined by the relations $E_{1\pm}-E_{00}=0$. For the case of $\Delta_{a}=\Delta_{b}=\Delta$, corresponding to the transitions $|\varepsilon_{00}\rangle\rightarrow |\varepsilon_{1\pm}\rangle$, the resonance relations are reduced to $\Delta=-(K\pm\sqrt{K^{2}+4J^{2}})/2$. For the parameters used in our simulations, the locations of the two peaks in $P_{|1,0\rangle}$ are given by $\Delta/\kappa=-16.1803$ and $6.1803$.
%%%%%%%%%%%%%%%%%%%%%%%
%%%%%%%%%%%%%%%%%%%%%%%

For the state occupation $P_{|2,0\rangle}$ (the red solid curve), there are four peaks, three of which correspond to the two-photon resonant transitions $\vert\varepsilon_{00}\rangle\rightarrow\vert\varepsilon_{2s}\rangle$ for $s=\pm,0$. The locations of these three main peaks are determined by the resonance conditions $E_{2s}-E_{00}=0$ for $s=\pm,0$. The two-photon resonance means that the system absorbs two driving photons and then transits from the state $\vert\varepsilon_{00}\rangle$ to one of the states $\vert\varepsilon_{2s}\rangle$ for $s=\pm,0$. In this physical process, the single-photon processes are nor resonant, but the energy change in the two-photon processes are conserved, i.e., the energy difference between the states $\vert\varepsilon_{2s}\rangle$ and the state $\vert\varepsilon_{00}\rangle$ equals to the energy of two driving photons. For the parameters used in our simulations, the locations of these three peaks in $P_{|2,0\rangle}$ are given by $\Delta/\kappa=-23.131$, $-7.576$, and $5.707$. In addition, there is a peak induced by the single photon resonance, and hence the location of this peak is the same as that of the peak in $P_{|1,0\rangle}$.

Besides the peaks, we can also see some dips in $P_{|1,0\rangle}$ and $P_{|2,0\rangle}$. Physically, these dips are caused by the quantum interference effect existing in the state transitions induced by the driving cavity. To clarify this point, we study the steady state of the system in the eigenstate representation and analyze the relationship between the bare state $|1,0\rangle$ ($|2,0\rangle$) and these eigenstates $|\varepsilon_{1\pm}\rangle$ ($|\varepsilon_{2s}\rangle$ for $s=\pm,0$). For clearly seeing the physical picture of quantum interference, we adopt the effective Hamiltonian method to simulate the evolution of the system. The state of the system can be expressed in the eigenstate representation as
\begin{equation}
\vert \psi(t)\rangle=D_{00}(t)\vert \varepsilon_{00}\rangle+D_{1+}(t)\vert\varepsilon_{1+}\rangle+D_{1-}(t)\vert \varepsilon_{1-}\rangle+D_{2+}(t)\vert \varepsilon_{2+}\rangle+D_{20}(t)\vert \varepsilon_{20}\rangle+D_{2-}(t)\vert \varepsilon_{2-}\rangle,\qquad  \label{stateeigenrepre}
\end{equation}
where $D_{00}(t)$, $D_{1\pm}(t)$, and $D_{2s}(t)$ are respectively the probability amplitudes of the eigenstates $\vert\varepsilon_{00}\rangle$, $\vert\varepsilon_{1\pm}\rangle$, and $\vert\varepsilon_{2s}\rangle$. Based on the Hamiltonian in Eq.~(\ref{Hnonheimite}) and the wave function in Eq.~(\ref{stateeigenrepre}), we can obtain the equations of motion for these probability amplitudes $D_{00}(t)$, $D_{1\pm}(t)$, and $D_{2s}(t)$. Using the perturbation method, the steady-state solution of these probability amplitudes can be obtained. Based on the state in Eq.~(\ref{stateeigenrepre}) and the steady-state solutions of these probability amplitudes, the state occupations $P_{|1,0\rangle}$ and $P_{|2,0\rangle}$ can be obtained as follows
\begin{equation}
P_{|1,0\rangle}=\left\vert D_{1+}C_{1,0}^{[1+]}\right\vert^{2}+\left\vert D_{1-}C_{1,0}^{[1-]}\right\vert^{2}+2\text{Re}\left[D_{1+}C_{1,0}^{[1+]}D_{1-}^{\ast}C_{1,0}^{[1-]\ast}\right],\label{rho44noninterf}
\end{equation}
and
\begin{align}
P_{|2,0\rangle}&=\left\vert D_{2+}C_{2,0}^{[2+] }\right\vert ^{2}+\left\vert
D_{2-}C_{2,0}^{[2-]}\right\vert ^{2}+\left\vert D_{20}C_{2,0}^{[20]}\right\vert^{2}\notag \\
&\quad+2\text{Re}\left[D_{2+}C_{2,0}^{[2+] }D_{2-}^{\ast}C_{2,0}^{[2-]\ast}
+D_{2+}C_{2,0}^{[2+]}D_{20}^{\ast}C_{2,0}^{[20]\ast}+D_{2-}C_{2,0}^{[2-]}D_{20}^{\ast}C_{2,0}^{[ 20]\ast}\right],\quad\label{rho66noninterf}
\end{align}
where ``$\text{Re}$'' gives the real part of the variable.

The first two terms of $P_{|1,0\rangle}$ in Eq.~(\ref{rho44noninterf}) and the first three terms of $P_{|2,0\rangle}$ in Eq.~(\ref{rho66noninterf}) are, respectively, the non-quantum-interference parts of the photon-number probabilities of $|1,0\rangle$ and $|2,0\rangle$, and the rest parts are induced by the quantum interference effect among the eigenstates in the same subspace of $N=0$, $1$, and $2$. To see the influence of the quantum interference on the photon blockade effect, we also show the non-quantum-interference parts (the dotted curves) of $P_{|1,0\rangle}$ and $P_{|2,0\rangle}$ as a reference in Fig.~\ref{Fig2}(a). Here we can confirm that for the state occupation $P_{|1,0\rangle}$ there are two main peaks which match the exact results. The dip confirmed by the exact result disappears in the non-quantum-interference result. This means that the dip in $P_{|1,0\rangle}$ is caused by the quantum interference effect between two transition paths  $|\varepsilon_{00}\rangle\rightarrow|\varepsilon_{1\pm}\rangle$. For the state occupation $P_{|2,0\rangle}$, we can also see a dip, which is induced by the quantum interference effect.

To clearly see the photon blockade effect in the linear cavity, in Fig.~\ref{Fig2}(b) we plot the equal-time second-order correlation function $g_{a}^{(2)}(0)$ as a function of the driving detuning $\Delta/\kappa$. Here, the blue dashed curve is plotted using the numerical solution of Eq.~(\ref{masteq}), while the red solid curve is based on the analytical solution given in Eq.~(\ref{gensolution}). To see the quantum interference effect in the conventional photon blockade regime, we also show the result corresponding to the non-quantum-interference parts using the grey dotted curve as a reference. We can see that the analytical result can match well with the numerical result, and that the non-quantum-interference evaluation can predict the location of the optimal driving detuning, but it cannot give the exact value of the correlation function $g_{a}^{(2)}(0)$. The locations of the dips ($d_{1}$ and $d_{2}$) of the correlation function $g_{a}^{(2)}(0)$ correspond to the single-photon resonance, namely the two peaks in $P_{|1,0\rangle}$. The locations of the peaks ($p_{1}$, $p_{2}$, and $p_{2}$) in $g_{a}^{(2)}(0)$ correspond to those of the peaks in $P_{|2,0\rangle}$. We find that the conventional photon blockade effect occurs in single-photon resonance case, i.e., $g_{a}^{(2)}(0)\ll1$. In addition, the photon blockade effect induced by quantum interference can be observed in the location of $d_{\text{int}}$.
%%%%%%%%%%%%%%%%%%%%%%%
\begin{figure}[tbp]
\center
\includegraphics[width=9cm]{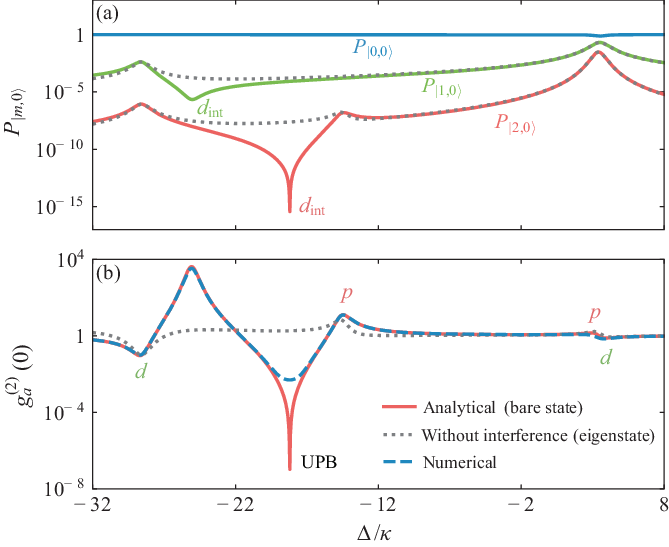}
\caption{(a) The state occupations $P_{|m,0\rangle}$ for the bare states $|m,0\rangle$ ($m=0,1,2$) versus the scaled driving detuning $\Delta/\kappa$. The colored solid curves and the gray dotted curves are plotted based on the exact results and the non-quantum-interference part of the exact results, respectively. (b) The correlation function $g_{a}^{(2)}(0)$ versus the driving detuning $\Delta/\kappa$. The blue dashed curve is plotted using the numerical solution of Eq.~(\ref{masteq}), while the red solid (gray dotted) curve corresponds to the exact (non-quantum-interference) analytical result. Here the dip corresponds to the unconventional photon blockade (UPB). Other parameters are $\kappa_{a}=\kappa_{b}=\kappa$, $\Omega/\kappa=0.3$, $K/\kappa=25.0916$, $J/\kappa=10$, and $\bar{n}_{a}=\bar{n}_{b}=0$.}
\label{Fig3}
\end{figure}
%%%%%%%%%%%%%%%%%%%%%%%

\subsection{Unconventional photon blockade}
The destructive quantum interference phenomenon between the two transition pathways is the physical origin of the unconventional photon blockade effect. Mathematically, to observe photon blockade effect in the linear cavity, it means that the two-photon probability in this cavity is suppressed. For the ideal case, the probability amplitude for the state $\vert 2,0\rangle$ is zero. Based on the solution~(\ref{stssoluprobaamplit}), the parameter condition for $C_{2,0}=0$ can be obtained as
\begin{align}
R[J,K,\Delta,\kappa]&=4J^{2}K+8K^{3}+28K^{2}\Delta+28K\Delta^{2}+8\Delta^{3}-7K\kappa^{2}-6\Delta\kappa^{2}=0,\notag\\
I[J,K,\Delta,\kappa]&=14K^{2}+28K\Delta+12\Delta^{2}-\kappa^{2}=0.\label{optimalcondition}
\end{align}
We point out that the condition in Eq.~(\ref{optimalcondition}) is obtained at $\omega_{a}=\omega_{b}$. If we choose $\omega_{a}=\omega_{b}+K$, then the corresponding condition is equal to the condition for the unconventional photon blockade in the two-nonlinear-cavity case~\cite{bamba2011Origin}. As shown in Ref.~\cite{bamba2011Origin}, the parameter condition is independent of the Kerr parameter of the first nonlinear cavity.

We can see that $R[J,K,\Delta,\kappa]$ is a function of $J^{2}$, which indicates that the unconventional photon blockades are the same for both attraction and repulsion interactions between the photons in the nonlinear cavity. In addition, we find that there exists the symmetric relations
\begin{equation}
R[J,-K,-\Delta,\kappa]=-R[J,K,\Delta,\kappa],\qquad
I[J,-K,-\Delta,\kappa]=I[J,K,\Delta,\kappa].
\end{equation}
Based on the symmetric relations and Eq.~(\ref{optimalcondition}), if $(K,\Delta)$ is a solution, then $(-K,-\Delta)$ is also a solution. As an example, we choose a moderate photon-hopping interaction strength $J/\kappa=10$, then the solutions of Eq.~(\ref{optimalcondition}) are:
\begin{subequations}
\label{upbsolution}
\begin{align}
&K/\kappa\approx\pm3.85\times10^{-3},\hspace{1 cm}\Delta/\kappa\approx\pm0.284;\label{upbsolutiona}\\
&K/\kappa\approx\pm 13.805i,\hspace{1.59 cm}\Delta/\kappa\approx\mp22.187i;\label{upbsolutionb}\\
&K/\kappa\approx\mp25.0916,\hspace{1.5 cm}\Delta/\kappa\approx\pm18.2054.\label{upbsolutionc}
\end{align}
\end{subequations}
%%%%%%%%%%%%%%%%%%%%%%%
\begin{figure}[tbp]
\center
\includegraphics[width=9cm]{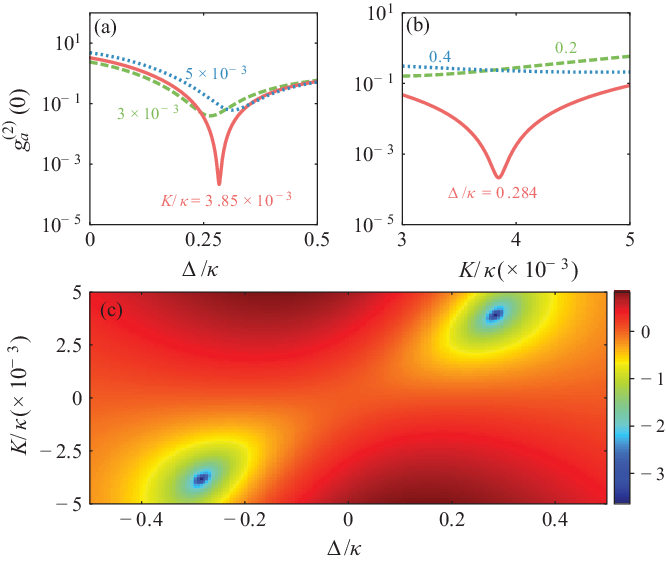}
\caption{(a) The correlation function $g_{a}^{(2)}(0)$ versus the driving detuning $\Delta/\kappa$ at various values $K/\kappa=(3, 3.85, 5)\times10^{-3}$. (b) The correlation function $g_{a}^{(2)}(0)$ versus the Kerr parameter $K/\kappa$ at various values $\Delta/\kappa=(0.2, 0.284, 0.4)$. (c) The correlation function $\log_{10}g_{a}^{(2)}(0)$ versus the driving detuning $\Delta/\kappa$ and the Kerr parameter $K/\kappa$.
Other parameters are $\kappa_{a}=\kappa_{b}=\kappa$, $J/\kappa=10$, $\Omega/\kappa=0.3$, and $\bar{n}_{a}=\bar{n}_{b}=0$.}
\label{Fig4}
\end{figure}
%%%%%%%%%%%%%%%%%%%%%%%

The two solutions in Eq.~(\ref{upbsolutiona}) correspond to the surprisingly-weak-Kerr nonlinearity case, which has been discussed in Ref.~\cite{bamba2011Origin}. The imaginary solutions in Eq.~(\ref{upbsolutionb}) should be discarded. The two solutions in Eq.~(\ref{upbsolutionc}) correspond to the strong-Kerr nonlinearity case, which means that unconventional photon blockade can also be obtained with a strong Kerr nonlinearity in this model. In order to further explain this phenomenon, in Fig.~\ref{Fig3}(a) we plot the state occupations $P_{|m,0\rangle}$ for the bare states $|m,0\rangle$ ($m=0,1,2$) versus the scaled driving detuning $\Delta/\kappa$ when the Kerr parameter $K/\kappa=25.0916$ and the photon-hopping strength $J/\kappa=10$. We also show the non-quantum-interference part (the dotted curves) of $P_{|1,0\rangle}$ and $P_{|2,0\rangle}$ as a reference. We can see some dips in $P_{|1,0\rangle}$ and $P_{|2,0\rangle}$, these dips are caused by the quantum interference effect. To clearly see the quantum interference effect on the photon blockade in the linear cavity, in Fig.~\ref{Fig3}(b) we plot the correlation function $g_{a}^{(2)}(0)$ versus the driving detuning $\Delta/\kappa$. We find that the unconventional photon blockade can be obtained at the detuning $\Delta/\kappa\approx-18.2054$ [i.e., the solutions in Eq.~(\ref{upbsolutionc})] corresponding to the dip in $P_{|2,0\rangle}$. This proves that the unconventional photon blockade is based on the destructive quantum interference effect and that the unconventional photon blockade also exists in a strong Kerr nonlinearity regime.

In order to confirm the optimal condition, we inspect the correlation function $g_{a}^{(2)}(0)$ when the system parameters take the values corresponding to the real solutions of Eq.~(\ref{optimalcondition}). In Fig.~\ref{Fig4}(a), we plot the correlation function $g_{a}^{(2)}(0)$ as a function of the driving detuning $\Delta/\kappa$ when $K/\kappa$ takes the values $(3,3.85,5)\times10^{-3}$. In Fig.~\ref{Fig4}(b), we plot the correlation function $g_{a}^{(2)}(0)$ as a function of the Kerr parameter $K/\kappa$ when $\Delta/\kappa$ takes the values $(0.2,0.284,0.4)$. As expected, the correlation function $g_{a}^{(2)}(0)$ shows a strong antibunching effect at $K/\kappa\approx 3.85\times10^{-3}$ and $\Delta/\kappa\approx0.284$, which indicates a clear signature of photon blockade in the linear cavity. The optimal parameter condition for unconventional photon blockade can be seen clearer by displaying the correlation function $g_{a}^{(2)}(0)$ as a function of the driving detuning $\Delta/\kappa$ and the Kerr parameter $K/\kappa$ for a given value of $J$. As illustrated in Fig.~\ref{Fig4}(c), there are two minimal-value points of $g_{a}^{(2)}(0)$ located at $\{K/\kappa\approx\pm3.85\times10^{-3},\Delta/\kappa\approx\pm0.284\}$, as analyzed by the symmetric relations.
%%%%%%%%%%%%%%%%%%%%%%%
\begin{figure}[tbp]
\center
\includegraphics[width=9cm]{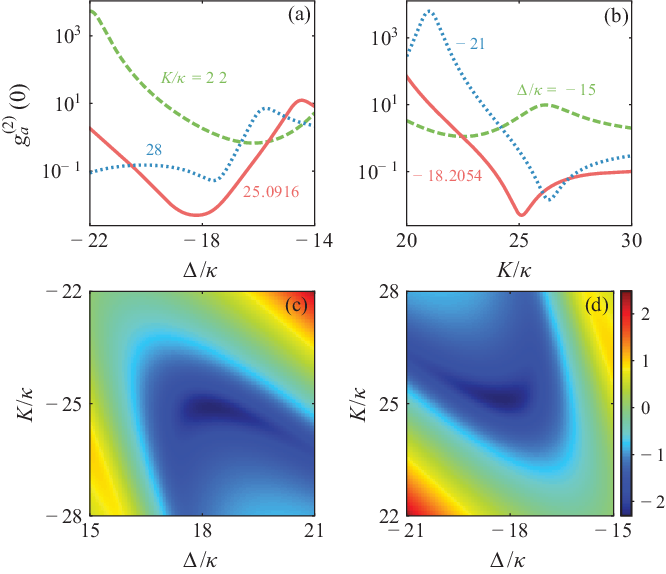}
\caption{(a) The correlation function $g_{a}^{(2)}(0)$ versus the driving detuning $\Delta/\kappa$ at various values $K/\kappa=(22,25.0916,28)$. (b) The correlation function $g_{a}^{(2)}(0)$ versus the Kerr parameter $K/\kappa$ at various values $\Delta/\kappa=(-15,-18.2054,-21)$. (c) and (d) The correlation function $\log_{10}g_{a}^{(2)}(0)$ versus the driving detuning $\Delta/\kappa$ and the Kerr parameter $K/\kappa$. Other parameters are $\kappa_{a}=\kappa_{b}=\kappa$, $J/\kappa=10$, $\Omega/\kappa=0.3$, and $\bar{n}_{a}=\bar{n}_{b}=0$.}
\label{Fig5}
\end{figure}
%%%%%%%%%%%%%%%%%%%%%%%

We also investigate the correlation function $g_{a}^{(2)}(0)$ corresponding to the real solutions $\{K/\kappa\approx\mp25.0916,\Delta/\kappa=\pm18.2054\}$. In Fig.~\ref{Fig5}(a), we plot the correlation function $g_{a}^{(2)}(0)$ as a function of the driving detuning $\Delta/\kappa$ when $K/\kappa$ takes the values $(22,25.0916,28)$. Similarly, in Fig.~\ref{Fig5}(b) we plot the correlation function $g^{(2)}(0)$ as a function of the Kerr parameter $K/\kappa$ when $\Delta/\kappa$ takes the values $(-15,-18.2054,-21)$. It is clear to see that the correlation function $g^{(2)}(0)$ shows a strong antibunching effect at $K/\kappa\approx25.0916$ and $\Delta/\kappa\approx-18.2054$. Here the locations of the two minimal value points are also determined by the symmetric relations.

Physically, the photon blockade effect corresponding to the above four real solutions $\{K/\kappa\approx\pm3.85\times10^{-3},\Delta/\kappa\approx\pm0.284\}$ and $\{K/\kappa\approx\mp25.0916,\Delta/\kappa\approx\pm18.2054\}$ can be explained using the destructive interference between two different paths of photon excitation. When the system parameters take the values corresponding to the real solutions of Eq.~(\ref{optimalcondition}), the probability amplitude $C_{2,0}=0$ is obtained, which indicates that the state $\vert 2,0\rangle$ will not be populated due to the destructive interference.

\section{The influence of the thermal excitation on the photon blockade effect}

\begin{figure}[tbp]
\center
\includegraphics[width=9cm]{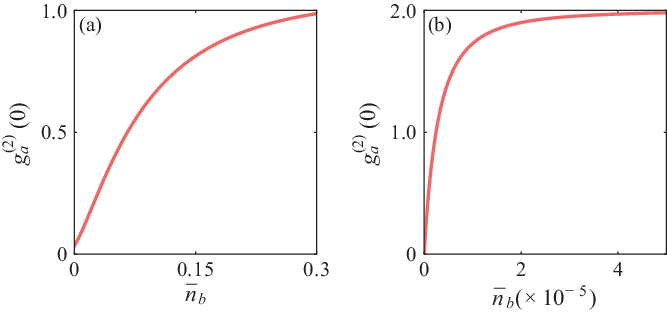}
\caption{The correlation function $g_{a}^{(2)}(0)$ as a function of the average thermal excitation numbers $\bar{n}_{b}$ in the conventional photon blockade case (a)
$\Delta=-(K+\protect\sqrt{K^{2}+4J^{2}})/2$ and $K/\kappa = 10$, and in the unconventional photon blockade case (b) $\Delta/\kappa=0.284$ and $K/\kappa=3.85\times10^{-3}$. Other parameters are $\kappa_{a}=\kappa_{b}=\kappa$, $J/\kappa=10$, $\bar{n}_{a}=0$, and $\Omega=0.3\kappa$.}
\label{Fig6}
\end{figure}

The above discussions focus on the zero-temperature environment case. For optical cavities, the thermal photon number is negligible, and hence our above assumption of the zero-temperature environments is reasonable. However, for other bosonic excitations such as phonon and magnon, it is needed to consider the influence of the thermal excitations on the photon blockade effects. Below, we will discuss the thermal bath case by considering a finite thermal excitations in the quantum master equation~(\ref{masteq}). In Fig.~\ref{Fig6}, we plot the correlation function $g_{a}^{(2)}(0)$ as a function of the thermal excitation number $\bar{n}_{b}$ in mode $b$. The value of $g_{a}^{(2)}(0)$ increases with the increase of the thermal photon number $\bar{n}_{b}$. It is clear that the thermal photons have a significant effect on both the conventional and unconventional photon blockade effects, and that the latter is more fragile against the thermal noise than the former. We can conclude that the thermal noise is fatal to the photon blockade effect in this coupled cavity system.

\section{Discussions}

We present some discussions on the experimental implementation of this scheme in some quantum optical systems such as coupled optical-cavity systems~\cite{gerace2009Quantumoptical}, coupled photon-magnon systems~\cite{wang2016Magnon,wang2018Bistability}, coupled microwave-resonator and phononic-resonator system~\cite{arrangoiz-arriola2018Coupling}, and coupled superconducting-resonator systems~\cite{kirchmair2013Observation,eichler2014QuantumLimited,magesan2018Effective}. We also present some analyses on the parameter conditions of these systems. In this model, there are three components: the self-Kerr interaction of mode $b$, the excitation hopping interaction between the two modes $a$ and $b$, the monochromatic driving of mode $a$. To implement this scheme, the candidate physical systems should have these three physical processes.

For coupled optical cavity system, the Kerr interaction can be implemented with a Kerr-type nonlinear cavity, and the Kerr parameter can enter the strong-coupling regime. A ratio $K/\kappa\sim10$ has been estimated in solid cavity-QED systems~\cite{gerace2009Quantumoptical}. The photon-hopping interaction between the two cavities can also be realized in the coupled cavity systems. Based on the detailed systems, the optical cavity could be various semiconductor microcavities and the Fabry-P\'{e}rot cavity. In these two cases, the driving on the cavity can be implemented.

For coupled photon-magnon systems, the two bosonic modes could be implemented with an electromagnetic mode in the superconducting resonator and a magnon mode of a yttrium iron garnet (YIG). Some recent experiments~\cite{wang2016Magnon,wang2018Bistability} reported that the Kerr interaction in the magnon mode can be implemented and the excitations hopping between photons and magnons can be realized. The cavity field driving can also be realized in superconducting resonator by introducing a microwave driving. In this system, the resonance frequency of the linear cavity (photon) mode is $\omega_{a}\approx2\pi\times10.1$ GHz, and the resonance frequency $\omega_{b}$ of the nonlinear bosonic (magnon) mode could range from several hundreds of megahertz to $28$ GHz. The coupling strength between the electromagnetic field and the magnonic mode is $J\approx 2\pi\times 42$ MHz. The decay rates of the cavity mode and the magnon mode are given by $\kappa_{a}\approx2\pi\times2.87$ MHz and $\kappa_{b}\approx2\pi\times24.3$ MHz. However, the magnitude of the self-Kerr interaction of the magnon mode is very small and hence the coupled photon-magnon model might be a possible candidate of the unconventional photon blockade effect in the weak-Kerr nonlinearity case.
%%%%%%%%%%%%%%%%%%%%%%%

For coupled superconducting resonator systems, the Kerr interaction can be implemented with the Josephson nonlinearity~\cite{kirchmair2013Observation}, and the photon-hopping interaction can be implemented with a capacitor or other superconducting elements~\cite{eichler2014QuantumLimited}. Recently, some experiments have been reported that similar systems have been realized in superconducting setups. In a recent reported experiment~\cite{kirchmair2013Observation}, the authors proposed a method to realize a highly coherent Kerr medium by coupling a superconducting vertical transmon qubit to two three dimensional waveguide cavities. In this system, the resonance frequency of the cavity is $\omega_{b}\approx2\pi\times9.2747$ GHz, the decay rate of the cavity mode is $\kappa_{b}\approx2\pi\times10$ kHz, and the Kerr parameters is $K\approx2\pi\times325$ kHz. Here we can see that the ratio $K/\kappa$ could be larger than $30$, and hence the strong Kerr nonlinearity in coupled cavity system is accessible. The coupling between two superconducting resonators can be realized through a capacitor~\cite{eichler2014QuantumLimited}, and the coupling strength between the two resonators could reach the order of megahertz. Therefore, the parameters used in the present paper is accessible with current experimental condition in superconducting quantum circuits. The monochromatic driving of the superconduction resonator can be realized through the microwave field, the driving frequency and driving amplitude can be controlled on demand. For current experimental condition, the resonance frequency of the superconducting resonators is of the order of $2\pi\times5$ - $10$ GHz, and the working temperature is about $15$ - $25$ mK. Based on these two parameter, we can estimate the thermal occupation number in the superconducting resonators. As an example, for $T\approx25$ mK, we have $n_{\textrm{th}}\approx4.9\times10^{-9}$ - $6.7\times10^{-5}$ when $\omega_{a}\approx2\pi \times5$ - $10$ GHz. In this case, the conventional photon blockade can be implemented in the coupled superconducting resonator system. To observe the unconventional photon blockade, the thermal noise can be suppressed by choosing a large resonance frequency and a low temperature work environment. Note that the unconventional photon blockade effect has recently been observed in a superconducting system~\cite{vaneph2018Observation}, which is described by the same physical model as that in this paper.

In this paper, we study the equal-time second-order correlation function of the cavity field. To observe the photon blockade effect in experiments, it requires the single-photon detection of the cavity photons in the steady state. This looses the technique requirements for the fast-resolution single-photon detection, which is needed in the measurement of time-delayed correlation function~\cite{bamba2011Origin}. In addition, the photon number detection can also be performed indirectly. For example, in a recent experiment~\cite{vaneph2018Observation}, the state in the resonator is assumed to be Gaussian, and then the second-order correlation function is obtained by measuring the two quadratures of the amplified field.

\section{Conclusion \label{conclusion}}

In conclusion, we studied the conventional and unconventional photon blockade effects in a coupled cavity system, which is formed by a linear cavity coupled to a Kerr-type nonlinear cavity via a photon-hopping interaction. Here the linear cavity is weakly driven by a monochromatic laser field. The photon blockade effect can be observed by calculating the equal-time second-order correlation function of the linear cavity field. We derived the optimal parameters required for the photon blockade by solving the non-Hermitian Schr\"{o}dinger equation with the decay in the system. We found that the photon blockade effect can be explained based on the anharmonicity of the eigenenergy spectrum and destructive quantum interference effect between different paths, respectively. In particular, we found that the quantum-interference-induced physical phenomenon also exists in the conventional photon blockade regime. In addition, the unconventional photon blockade can occur in both the weak and the strong Kerr-nonlinearity cases in this model. Some discussions on the experimental feasibility of this scheme are presented.

\section*{Funding}
J.-Q.L. is supported in part by National Natural Science Foundation of China (Grants No.~11822501, No.~11774087, and No.~11935006), Natural Science Foundation of Hunan Province, China (Grant No.~2017JJ1021), and Hunan Science and Technology Plan Project (Grant No.~2017XK2018). D.-G.L. is supported in part by Hunan Provincial Postgraduate Research and Innovation project (Grant No.~CX2018B290).

\section*{Acknowledgments}
J.-Q.L. thanks Prof. Xun-Wei Xu for helpful discussions.

\section*{Disclosures}
The authors declare no conflicts of interest.

%\bibliography{RefCCS}

\end{document}